\documentclass[12pt]{article}
\pdfoutput=1
\usepackage{amsmath}
\usepackage{graphics,graphicx,xcolor}

\def\be{\begin{equation}}
\def\ee{\end{equation}}
\def\arr{\begin{array}{rll}}
\def\ea{\end{array}}
\def\bea{\begin{eqnarray}}
\def\eea{\end{eqnarray}}

\def\N2{$N{=}2$}

\def\>{\rangle}
\def\<{\langle}
\def\+{\dagger}
\def\={\ =\ }

\def\bal{\begin{aligned}}
\def\eal{\end{aligned}}

\begin{document}
\begin{titlepage}
\setcounter{page}{0}
\begin{center}
{\LARGE\bf  Remarks on Galilean electromagnetism}\\
\vskip 1.2cm
\textrm{\Large Anton Galajinsky \ }
\vskip 0.7cm
{\it
Tomsk Polytechnic University, 634050 Tomsk, Lenin Ave. 30, Russia} \\
\vskip 0.2cm
{\it
Tomsk State University of Control Systems and Radioelectronics,\\
Lenin ave. 40, 634050 Tomsk, Russia} \\

\vskip 0.2cm
{e-mail: galajin@tpu.ru}
\vskip 0.5cm
\end{center}

\begin{abstract} \noindent
Symmetries of the nonrelativistic electrodynamics are studied.
It is shown that equations describing the Galilean electromagnetism 
in the presence of sources
hold invariant under 
the $\ell$--conformal Galilei group for an arbitrary (half)integer parameter $\ell$. 
The group contains a transformation which links an inertial 
frame of refe\-rence to that moving with a constant acceleration of order up to $2\ell-1$, thus
pointing at
potential dynamical instability of the system.    

\end{abstract}

\vspace{0.5cm}

\end{titlepage}
\renewcommand{\thefootnote}{\arabic{footnote}}
\setcounter{footnote}0

\noindent
{\bf 1. Introduction}\\

Extensive recent studies of the nonrelativistic AdS/CFT--correspondence extended
the holographic dictionary to cover strongly coupled condensed matter systems 
(see \cite{NS} and 
references therein). As far as symmetries are concerned, the key ingredient here is a 
finite--dimensional conformal extension of the Galilei group \cite{Henkel,NOR}.

Generators of the corresponding Lie algebra include
(temporal) translation, dilatation, and special conformal transformation, 
which form $sl(2,R) \sim so(2,1)$ subalgebra, as well as spatial translations, Galilei boosts 
and higher order constant accelerations.\footnote{The algebra also involves generators of
spatial rotations, which in what follows will be disregarded.}  The associated structure relations
involve an arbitrary (half)integer parameter $\ell$, which specifies $2\ell-1$ acceleration 
generators at hand  and gives the name to the algebra \cite{Henkel,NOR}. 
In the literature, the reciprocal of $\ell$ is called the rational dynamical exponent. 
It is also customary to
refer to the instances of $\ell=\frac 12$ and $\ell=1$
as the Schr\"odinger algebra and the conformal Galilei
algebra, respectively (for a recent review see \cite{DHHRZ}).

The study of dynamical realizations of the $\ell$–conformal Galilei group 
revealed an interesting peculiarity. Because the number of functionally independent 
integrals of
motion needed to integrate a differential equation correlates with its order, 
such systems in
general involve higher derivative terms (see e.g. \cite{DH1}--\cite{TS1} and 
references therein\footnote{The only known example in the literature of a second 
order system, for which all constants of motion are functionally
independent, was built in \cite{CG} by making recourse to geodesics on a Ricci--flat 
spacetime with the $\ell$--conformal Galilei isometry group.}), the order of the 
highest derivative typically 
being $2\ell+1$.  

A particularly interesting example in the chain of models accommodating 
the $\ell$--conformal Galilei symmetry is the Galilean electromagnetism \cite{BLL}.
As was demonstrated in \cite{BBM}, the corresponding equations of motion in the 
absence of sources hold invariant under the 
infinite dimensional extension of the $\ell=1$ conformal Galilei group, which was 
called in \cite{BBM}
the infinite dimensional Galilean conformal group.

The goal of this paper is twofold. Firstly, it is shown that equations describing 
the Galilean 
electromagnetism 
in the presence of sources
actually hold invariant under the $\ell$--conformal Galilei group for 
an arbitrary (half)integer parameter $\ell$. Secondly, it is emphasized that the system
may potentially be plagued with dynamical instabilities,
because the group contains a transformation which links an inertial 
frame of reference to that moving with a constant acceleration of order up to $2\ell-1$ (see Sect. 5).
 
The work is organized as follows. In the next section, the electric 
and magnetic nonrela\-tivistic limits
of the Maxwell equations are outlined. In Sect. 3, the structure of 
the $\ell$--conformal Galilei group is briefly reminded. In Sect. 4, it is demonstrated 
that equations describing the Galilean electromagnetism 
in the presence of sources
hold invariant under 
the $\ell$--conformal Galilei group for an arbitrary (half)integer parameter $\ell$. Physical implications of
such symmetry are discussed in the concluding Sect. 5.

Throughout the paper, summation over repeated indices is understood unless otherwise stated.

\vspace{0.5cm}

\noindent
{\bf 2. Nonrelativistic limits of the Maxwell equations}\\

When exploring nonrelativistic limits of the Maxwell equations\footnote{In this section, 
we use notations similar to those in \cite{LL}. In particular, 
$x^\mu=(c t,x^i)$, $\partial_\mu=\frac{\partial}{\partial x^\mu}$, 
$\eta_{\mu\nu}={\mbox{diag}}\left(+,-,-,-\right)$, $A^\mu=(\varphi,{\bf A})$,
$A_\mu=(\varphi,-{\bf A})$, $j^\mu=(c \rho, {\bf j})$, $F_{\mu\nu}=\partial_\mu A_\nu-\partial_\nu A_\mu$, 
$F_{0i}=-\partial_i \varphi-\frac{1}{c} \partial_t A_i:=E_i$, 
$F_{ij}=-\partial_i A_j+\partial_j A_i:=-\epsilon_{ijk} H_k$, with $i,j,k=1,2,3$, where $c$ is the speed of light.} 
\be\label{ME}
\partial_\mu F^{\nu\mu}=-\frac{4\pi}{c} j^\nu, \qquad \partial_\mu  j^\mu=0,
\nonumber
\ee
it is customary to rescale the scalar and vector potentials 
$\varphi \to \alpha(c) \varphi$, ${\bf A} \to \beta(c) {\bf A}$, as well as the charge 
density $\rho \to \gamma(c) \rho$ and the current density ${\bf j} \to \sigma(c) {\bf j}$, and then 
choose the factors $\alpha(c),\beta(c),\gamma(c),\sigma(c)$ in such a way that the resulting equations are 
nonsingular in the limit $c \to \infty$. 

In general, more than one consistent nonrelativistic 
limit is possible and 
additional insight comes from the Lorentz boost transformation
\bea\label{LT}
&&
\varphi=\frac{\varphi'+\frac{v}{c}A'^1}{\sqrt{1-{\left(\frac{v}{c}\right)}^2}}, \qquad 
A^1=\frac{A'^1+\frac{v}{c} \varphi'}{\sqrt{1-{\left(\frac{v}{c}\right)}^2}}, \qquad 
A^2=A'^2, \qquad A^3=A'^3,
\nonumber\\[2pt]
&&
j^0=\frac{j'^0+\frac{v}{c} j'^1}{\sqrt{1-{\left(\frac{v}{c}\right)}^2}}, \qquad 
j^1=\frac{j'^1+\frac{v}{c} j'^0}{\sqrt{1-{\left(\frac{v}{c}\right)}^2}}, \qquad 
j^2=j'^2, \qquad j^3=j'^3.
\eea
It suggests a particular rescaling ${\bf A} \to \frac{1}{c} {\bf A}$, 
with other fields unchanged, which guarantees that 
both the resulting transformation\footnote{The nonrelativistic limit of the Lorentz boost transformation
acting in space--time yields the conventional Galilei boost $t=t'$, $x^i=x'^i+v^i t'$.} 
\bea\label{TRel}
\varphi=\varphi', \qquad {\bf A}={\bf A}'+{\bf v} \varphi', \qquad
\rho=\rho', \qquad {\bf j}={\bf j}'+{\bf v} \rho',
\eea
and the equations of motion 
\bea\label{EL}
\mbox{div} {\bf E}=4 \pi \rho, \quad ~ \mbox{rot} {\bf E}=0, \quad ~
\mbox{rot} {\bf H}=\frac{\partial {\bf E}}{\partial t}+4 \pi {\bf j}, \quad ~ \mbox{div} {\bf H}=0,
\quad ~
\frac{\partial\rho}{\partial t}+\mbox{div} {\bf j}=0,  
\eea
where ${\bf E}=-\mbox{grad} \varphi$, ${\bf H}=\mbox{rot} {\bf A}$, 
are nonsingular after taking the limit $c \to \infty$. In the literature,
this instance is known as the electric limit of the Maxwell equations \cite{BLL}.

Alternatively, one can choose ${\bf A} \to c {\bf A}$, 
${\bf j} \to c^2 {\bf j}$ (with $\varphi$ and $\rho$ unchanged),  $c \to \infty$, which result in
\bea\label{TRmag}
&&
\varphi=\varphi'+({\bf v},{\bf A}'), \qquad {\bf A}={\bf A}', \qquad
\rho=\rho'+({\bf v},{\bf j}'), \qquad {\bf j}={\bf j}',
\eea
and
\bea\label{ML}
&&
\mbox{div} {\bf E}=4 \pi \rho, \qquad \mbox{rot} {\bf E}=-\frac{\partial {\bf H}}{\partial t}, 
\qquad
\mbox{rot} {\bf H}=4 \pi {\bf j}, \qquad \mbox{div} {\bf H}=0,
\qquad
\mbox{div} {\bf j}=0,
\eea
where ${\bf E}=-\mbox{grad} \varphi-\frac{\partial {\bf A}}{\partial t}$, ${\bf H}=\mbox{rot} 
{\bf A}$. 
The latter case is referred to as the magnetic limit of the Maxwell equations \cite{BLL}.

Note that in the original work \cite{BLL} the need to introduce the nonrelativistic transformations 
(\ref{TRel}) or (\ref{TRmag}) was motivated by the fact that the conventional low velocity limit of (\ref{LT}) does not obey the 
group composition law, if transformation parameters are finite.

Below we demonstrate that (\ref{EL}) and (\ref{ML}) also hold invariant under the action of the 
$\ell$--conformal Galilei group \cite{Henkel,NOR} for an arbitrary 
(half)integer parameter $\ell$. 

\vspace{0.5cm}

\noindent
{\bf 3. The $\ell$--conformal Galilei group}\\

The $\ell$--conformal Galilei group \cite{Henkel,NOR} includes
(temporal) translation, dilatation, and special
conformal transformation, which form $SL(2,R)$ subgroup, as well as 
spatial translations, Galilei boosts and higher order constant accelerations.
Structure relations of the corresponding Lie algebra involve an arbitrary (half)integer 
parameter $\ell$, which specifies $2\ell-1$ acceleration generators 
at hand. 

In a nonrelativistic spacetime parametrized by a temporal variable $t$ and spatial 
coordinates $x_i$, $i=1,2,3$,
the $SL(2,R)$--transformations are realized as follows
\be\label{sl2}
t'=\frac{\alpha t+\beta}{\gamma t+\delta}, \qquad
x'_i={\left(\frac{\partial t'}{\partial t} \right)}^\ell x_i,
\ee
where $\alpha$, $\beta$, $\gamma$, $\delta$ are real parameters obeying 
$\alpha \delta-\beta \gamma=1$ and $\ell$ is a (half)integer number, while 
(no sum over $n$ below)
\be\label{acs}
t'=t, \qquad x'_i=x_i+a^{(n)}_i t^n,
\ee
with $n=0,\dots, 2\ell$, specify the spatial translation $\left(a^{(0)}_i\right)$, 
the Galilei boost $\left(a^{(1)}_i \right)$ and
higher order constant accelerations $\left( a^{(2)}_i, 
\dots, a^{(2\ell)}_i \right)$. 

Generators of the corresponding infinitesimal transformations 
\be\label{gener}
H=\frac{\partial}{\partial t}, \qquad 
D=t \frac{\partial}{\partial t}+\ell x_i \frac{\partial}{\partial x_i}, \qquad
K=t^2 \frac{\partial}{\partial t}+2 \ell t x_i \frac{\partial}{\partial x_i},
\qquad
C^{(n)}_i=t^n  \frac{\partial}{\partial x_i}
\nonumber
\ee
obey the structure relations of the $\ell$--conformal Galilei algebra \cite{Henkel,NOR}~\footnote{In arbitrary spatial dimension and 
for half-integer $\ell$ the algebra admits one central extension. As far as dynamical realizations are concerned, the central charge is usually 
interpreted as the mass of a (higher derivative) particle. In two spatial dimensions and for integer $\ell$ another central extension is allowed, 
which is usually attributed to the noncommutativity of space or external electric/magnetic fields present. For a detailed discussion of the central charges and further references to the original literature see e.g. \cite{MT,GM1}.}
\begin{align}\label{algebra}
&
[H,D]=H, &&  [H,K]=2D, && [D,K]=K,
\nonumber\\[2pt]
&
[H,C^{(n)}_i]=n C^{(n-1)}_i, && [D,C^{(n)}_i]=(n-l) C^{(n)}_i, 
&& [K,C^{(n)}_i]=(n-2l) C^{(n+1)}_i,
\end{align}
where $n=0,\dots, 2\ell$, $i=1,2,3$,
and $\ell$ is a (half)integer number. The fact that $\ell$ is (half)integer
guarantees that the algebra is finite--dimensional (mind the last bracket in (\ref{algebra})).

If one disregards the $SL(2,R)$--transformations, the range of index $n$ carried by
the vector generators $C^{(n)}_i$ is no longer linked to $\ell$. Taking $n$ to be
a natural number greater than one, one obtains a transformation which links an inertial 
frame of reference to that moving with a constant acceleration of order $n-1$.

\vspace{0.5cm}

\noindent
{\bf 4. The $\ell$--conformal Galilei symmetry of the Galilean electromagnetism}\\

Let us demonstrate that the $\ell$--conformal Galilei group is a symmetry of eqs. (\ref{EL}).
 
Transformation law of the charge density $\rho(t,x)$ 
is obtained by fixing a value of the temporal variable 
$t$ and demanding the charge contained within a three--dimensional volume element $V$ to be 
invariant under (\ref{sl2}), (\ref{acs})
\be
\int_{V'} d^3 x' \rho' (t',x')=\int_{V} d^3 x \rho(t,x).
\nonumber
\ee
This gives
\be\label{trr}
\rho(t,x)= {\left(\frac{\partial t'}{\partial t} \right)}^{3 \ell} \rho' (t',x')
\nonumber
\ee
for the $SL(2,R)$--transformation and
\be\label{trr1}
\rho(t,x)=\rho' (t',x')
\nonumber
\ee
for the transformation involving parameters $a^{(n)}_i$. 

Taking into account the identities 
\be\label{id}
\frac{\partial}{\partial t}=\left(\frac{\partial t'}{\partial t}\right) \frac{\partial}{\partial t'}+\left(\frac{\partial x'_i}{\partial t} \right) \frac{\partial}{\partial x'_i},
\qquad \frac{\partial}{\partial x_i}=\left( \frac{\partial t'}{\partial x_i} \right) \frac{\partial}{\partial t'}+\left(\frac{\partial x'_j}{\partial x_i} \right)\frac{\partial}{\partial x'_j},
\ee
and demanding the invariance of the continuity equation 
$\frac{\partial\rho}{\partial t}+\mbox{div} {\bf j}=0$,
one can then establish how the current density is transformed
\bea
&&
j_i (t,x)={\left(\frac{\partial t'}{\partial t} \right)}^{2 \ell+1} j'_i (t',x')-\ell 
{\left(\frac{\partial t'}{\partial t} \right)}^{2 \ell-1} \left(\frac{\partial^2 t'}{\partial t^2} \right) 
x'_i \rho' (t',x'),
\nonumber\\[2pt]
&&
j_i (t,x)= j'_i (t',x')-n t^{n-1} \rho' (t',x') a^{(n)}_i.
\eea

Given the transformation laws of the charge density above, it is straightforward to verify that
$\mbox{div} {\bf E}=4 \pi \rho$ holds invariant provided
\be\label{TRE}
E_i (t,x)={\left(\frac{\partial t'}{\partial t} \right)}^{2 \ell} E'_i (t',x'),
\qquad E_i (t,x)=E'_i (t',x'),
\ee
where the former relation specifies the $SL(2,R)$--transformation of the electric field strength,
while the latter 
links 
to the transformations involving $a^{(n)}_i$. Taking into account the identities (\ref{id}) and 
applying (\ref{sl2}), (\ref{acs}), (\ref{TRE}) to 
$\mbox{rot} {\bf E}=0$, one can establish the invariance of the latter equation as well.

Finally, demanding the invariance of 
$\mbox{rot} {\bf H}=\frac{\partial {\bf E}}{\partial t}+4 \pi {\bf j}$ and
$\mbox{div} {\bf H}=0$, one can determine how the magnetic 
field strength transforms under the $\ell$--conformal Galilei group (no sum over $n$ below)
\bea
&&
H_i (t,x)={\left(\frac{\partial t'}{\partial t} \right)}^{\ell+1} H'_i (t',x')-\ell 
{\left(\frac{\partial t'}{\partial t} \right)}^{\ell-1} 
\left(\frac{\partial^2 t'}{\partial t^2} \right) \epsilon_{ijk}
x'_j E'_k (t',x'),
\nonumber\\[2pt]
&&
H_i (t,x)=H'_i (t',x')-n t^{n-1} \epsilon_{ijk}
a^{(n)}_j E'_k (t',x'),
\eea
where $\epsilon_{ijk}$ is the totally antisymmetric symbol with $\epsilon_{123}=1$.

Note that, when verifying the invariance of the equations of motion for the magnetic 
field strength, the conditions $\mbox{rot} {\bf E}=0$, 
$\mbox{div} {\bf E}=4 \pi \rho$ have been used as well. To put it in other words, the 
full system of equations (\ref{EL}) holds invariant rather than each individual equation.
 
The magnetic limit of the Maxwell equations can be analyzed in a similar fashion.
It is straightforward to verify that eqs. (\ref{ML}) hold invariant under the $SL(2,R)$--transformations 
\bea
&&
\rho (t,x)={\left(\frac{\partial t'}{\partial t} \right)}^{2\ell+1} \rho' (t',x')
+\frac{\ell}{4 \pi}
{\left(\frac{\partial t'}{\partial t} \right)}^{2\ell-1}
\left(\frac{\partial^2 t'}{\partial t^2} \right) \epsilon_{ijk}
x'_j \frac{\partial H'_k (t',x')}{\partial x'_i}, 
\nonumber\\[2pt]
&&
j_i (t,x)={\left(\frac{\partial t'}{\partial t} \right)}^{3\ell} j'_i (t',x'),
\nonumber\\[2pt]
&&
E_i (t,x)={\left(\frac{\partial t'}{\partial t} \right)}^{\ell+1} E'_i (t',x')+ 
\ell {\left(\frac{\partial t'}{\partial t} \right)}^{\ell-1} 
\left(\frac{\partial^2 t'}{\partial t^2} \right) \epsilon_{ijk}
x'_j H'_k (t',x'), 
\nonumber\\[2pt]
&&
H_i (t,x)={\left(\frac{\partial t'}{\partial t} \right)}^{2\ell} 
H'_i (t',x'),
\nonumber
\eea
as well as under the transformations involving vector parameters $a^{(n)}_i$ 
(no sum over $n$ below)
\bea\label{acs1}
&&
\rho (t,x)=\rho' (t',x')+\frac{1}{4 \pi} n t^{n-1}
 \epsilon_{ijk}
a^{(n)}_j \frac{\partial H'_k (t',x')}{\partial x'_i} , \qquad
j_i (t,x)=j'_i (t',x'),
\nonumber\\[2pt]
&&
E_i (t,x)=E'_i (t',x')+n t^{n-1} \epsilon_{ijk}
a^{(n)}_j H'_k (t',x'), \qquad H_i (t,x)=H'_i (t',x').
\eea

Somewhat surprisingly, the invariance of the charge contained within a three--dimensional volume element
$\int_{V} d^3 x \rho(t,x)$
under the $SL(2,R)$--transformations then requires $\ell=1$ and it also
implies that both $H_i (t,x)$ and $\epsilon_{ijk} x_j H_k (t,x)$ should vanish at spatial infinity. 

When verifying the symmetries of eqs. (\ref{EL}) and (\ref{ML}) above the identities
\be
\epsilon_{ijk} \epsilon_{lpk}=\delta_{il}\delta_{jp}-\delta_{ip}\delta_{jl}, \qquad
\epsilon_{ijk} \epsilon_{ijp}=2 \delta_{kp}
\nonumber
\ee
proved useful.

It is important to stress that, discarding the $SL(2,R)$--transformations, both the electric and 
magnetic limit equations 
and the charge
$\int_{V} d^3 x \rho(t,x)$ still hold invariant under
a transformation which links an inertial 
frame of reference to that moving with a constant acceleration of order up to $2\ell-1$.
This latter fact will prove important 
for our discussion in the next section.

\vspace{0.5cm}

\noindent
{\bf 5. Discussion}\\

Due credit given to the elegant equations (\ref{EL}) and (\ref{ML}),
some of their properties deserve a critical discussion. For definiteness, 
below we focus on the electric limit. The magnetic limit can be considered likewise.

First, 
because $\frac{\partial {\bf H}}{\partial t}$ does not contribute to (\ref{EL}), it appears
problematic to
regard (\ref{EL}) as a fully consistent set of {\it evolutionary equations }
to determine ${\bf E}$ and ${\bf H}$. 

Second, for a similar reason the magnetic field strength ${\bf H}$ does not contribute
to the conserved energy 
\be\label{CE}
\frac{1}{8\pi} \int d^3 x ({\bf E},{\bf E}),
\ee
which characterizes the system in the absence of sources.\footnote{It is assumed 
here that $\epsilon_{ijk} E_j H_k$ vanishes at spatial infinity.}

Third, 
as was shown in the previous section,
eqs. (\ref{EL}) hold invariant under a transformation which links an inertial 
frame of reference to that moving with a constant acceleration of order up to $2\ell-1$.
In particular,
given a solution to (\ref{EL}) describing constant homogeneous electric and magnetic fields in a vacuum
\be\label{PR}
E_i (t,x)=\alpha_i, \qquad H_i (t,x)=\beta_i, \qquad \rho (t,x)=0, \qquad j_i (t,x)=0, 
\ee
where $\alpha_i$, $\beta_i$ are arbitrary vectors, the transformations yield another solutions to the same equations
\be\label{PR1}
E_i (t,x)=\alpha_i, \qquad H_i (t,x)=\beta_i+t^{n-1} \gamma_i, \qquad \rho (t,x)=0, \qquad j_i (t,x)=0, 
\ee
where $\gamma_i$ is an arbitrary vector, $t$ is the temporal variable, and 
$n=1,\dots,2\ell$.
Furthermore, $t^{n-1}$ in (\ref{PR1}) can actually be replaced with an arbitrary function $f(t)$, 
for example, $e^t$ (see below). 
An exponential growth over time of a magnetic field strength in the direction specified 
by a vector $\gamma_i$ in an empty space 
is certainly an unphysical phenomenon pointing at potential dynamical instability of the system. 
The fact that the magnetic field strength can grow unbounded and yet it does not contribute to 
the conserved energy (\ref{CE}) 
calls for a better understanding of count of physical degrees of freedom 
in phase space. A natural way to proceed is to implement the constrained dynamics analysis for the 
Lagrangian formulations in \cite{FHHO,BBM,HHO,BGK,FNG}. The key point is to understand whether there 
is a propagating degree of freedom supporting the magnetic field strength.\footnote{The author thanks an anonymous reviewer for
calling his attention to this fact.}

Fourth, consider the electric limit equations in the Lorentz gauge\footnote{Note that
the Lorentz gauge $\frac{\partial \varphi}{\partial t}+\mbox{div} {\bf A}=0$ holds invariant under the 
Galilei boost transformation (\ref{TRel}) accompanied by $t=t'$, $x^i=x'^i+v^i t'$.} 
\bea
\Delta \varphi=4 \pi \rho, \qquad \Delta {\bf A}= 4 \pi {\bf j}, 
\qquad \frac{\partial \varphi}{\partial t}+\mbox{div} {\bf A}=0, \qquad 
\frac{\partial\rho}{\partial t}+\mbox{div} {\bf j}=0,
\eea
where $\Delta$ is the Laplacian and we redefined $\varphi \to -\varphi $,
$\bf{A} \to -\bf{A}$. In the absence of sources, the equations admit a particular solution
\be
{\bf A}=\frac{ e^t {\bf a}}{\sqrt{({\bf x},{\bf x})}}, \qquad 
\varphi=\frac{ e^t ({\bf a},{\bf x})}{{({\bf x},{\bf x})}^{\frac 32}},
\ee
where ${\bf a}$ is a constant vector, meaning that 
the force exerted upon a charged particle moving on such a background
$m \ddot{\bf x}=e {\bf E}$, with
${\bf E}=\mbox{grad} \varphi$,
grows exponentially with time.\footnote{By symmetry argument,
in the electric limit case only the electric field strength contributes to the Lorentz force
\cite{BLL}.}  Again, this seems to indicate a potential dynamical instability. 

Fifth, in the literature on the Galilean electromagnetism, it is widely
underestimated that the Galilei boost (\ref{TRel}) is actually a particular
instance of a more general symmetry
\begin{align}\label{gen}
&
t'=t, && x'_i=x_i+\alpha_i (t), 
\nonumber\\[2pt]
&
\rho (t,x)=\rho' (t',x'), && j_i (t,x)=j'_i (t',x')-\dot{\alpha}_i (t) \rho' (t',x'), 
\nonumber\\[2pt]
&
E_i (t,x)=E'_i (t',x'), && H_i (t,x)=H'_i (t',x')-\epsilon_{ijk} \dot{\alpha}_j (t) E'_k (t',x'),
\end{align}
in which $\alpha_i (t)$ is an arbitrary differentiable function of time
and the dot designates the derivative with respect to $t$. The conventional Galilei boost emerges at 
$\alpha_i (t)=-t v_i$. In the absence of sources, the transformation (\ref{gen})
was first revealed in \cite{FHHO}, where it was also argued that the vacuum equations
of motion are time reparametrization invariant and one can 
perform a time--dependent spatial dilatations.
In the presence of sources, one can establish the following symmetry transformation rules
\begin{align}
&
t'=f(t), && x'_i=g(t) x_i, 
\nonumber\\[2pt]
&
\rho(t,x)=g^3 (t) \rho'(t',x'), &&
j_i (t,x)=g^2 (t) \dot{f} (t) j'_i (t',x')-g (t) \dot{g} (t) x'_i \rho'(t',x'),
\nonumber\\[2pt]
&
E_i (t,x)=g^2 (t) E'_i (t',x'), && H_i (t,x)=g(t) \dot{f}(t) H'_i (t',x')-
\dot{g}(t) \epsilon_{ijk} x'_j E'_k (t',x'),
\nonumber
\end{align}
where $f(t)$ and $g(t)$ are arbitrary differentiable functions of time.
The $\ell$--conformal Galilei symmetry 
described above thus appears to be a finite
dimensional subgroup of the larger infinite dimensional symmetry group.

Finally, in order to resolve some physical issues, like the problem with the definition of the conserved energy above, in the original work \cite{BLL}
it was suggested to double the number of fields and consider eqs. (\ref{EL}) and (\ref{ML}) 
simultaneously by assigning the subscript $"e"$ to the fields entering (\ref{EL}) and $"m"$ to those 
contributing to (\ref{ML}). The Lorentz force was modified accordingly by allowing 
mixed contributions. Yet, the problem with potentially unstable solutions
persists to the extended framework.

Turning to possible further developments, it would be interesting to explore whether the $\ell$-conformal Galilei symmetry 
discussed in this work can be realized at the level of action functionals studied in \cite{FHHO,BBM,HHO,BGK,FNG}.

\vspace{0.5cm}

\noindent{\bf Acknowledgements}\\

\noindent
This work was supported by the RF Ministry of Science and Higher Education 
under the project FEWM-2026-0010.

\end{document}